\documentclass[preprint,showpacs,preprintnumbers,amsmath,amssymb, graphicx,  graphics, epsfig]{revtex4}
\input{epsf}
\usepackage{amssymb}
\usepackage{bm}

\begin{document}

\def\be{\begin{equation}}
\def\ee{\end{equation}}
\def\bdm{\begin{displaymath}}
\def\edm{\end{displaymath}}
\def\erfc{\hbox{erfc }}
\def\vpa{v_{\parallel }}
\def\vper{v_{\perp }}
\def\Omm{\Omega }
\def\ppa{p_{\parallel }}
\def\pper{p_{\perp }}
\def\ppv{\vec{p}}
\def\kkv{\vec{k}}
\def\omm{\omega }

\title{\bf Covariant kinetic theory of nonlinear plasma waves interaction}
\author{M. Lazar}
\altaffiliation[Also at ]{"Alexandru Ioan Cuza" University, Faculty of Physics, 6600 Iasi, Romania}
\author{R. Schlickeiser}
\email{rsch@tp4.rub.de}
\affiliation{Institut f\"ur Theoretische Physik, Lehrstuhl IV:
Weltraum- und Astrophysik, Ruhr-Universit\"at Bochum,
D-44780 Bochum, Germany}
\altaffiliation[Also at ]{Centre for Plasma Science and Astrophysics,
Ruhr-Universit\"at Bochum, D-44780 Bochum, Germany}




\begin{abstract}
A rigorous and most general covariant kinetic formalism is
developed to study the nonlinear waves interaction in relativistic
Vlasov plasmas. The typical nonlinear plasma reaction is a
nonlinear current measured by the nonlinear plasma conductivity,
and these quantities are derived here on the basis of relativistic
Vlasov-Maxwell equations. Knowing the nonlinear plasma
conductivity allows us to determine all plasma modes nonlinearly
excited in plasma. The general covariant form of nonlinear
conductivity is provided first for any value of plasma temperature
and for the whole complex frequency plane by a correct analytical
continuation. The further analysis is restricted to a correct
relativistic particle distribution which is vanishing for particle
speeds greater than speed of light. In the limit of
nonrelativistic plasma temperatures the covariant nonlinear
conductivity is significantly different from the standard
noncovariant nonrelativistic results which are reached only in the
formal limit of an infinitely large speed of light $c \to \infty$.
\end{abstract}

\pacs{52.25.Dg; 52.27.Aj; 52.27.Ny; 52.35.Mw; 52.38.Bv}
\maketitle

\section{INTRODUCTION}

Under various sources of free energy plasma becomes a high nonlinear dispersive medium.
Numerous nonlinear wave processes occur in both laboratory and space plasmas [\cite{S86}]: 
e.g. in laser plasma interaction the high intensity pump waves stimulate the parametric instabilities and
the nonlinear collisionless damping of plasma waves; or the shock structures arising in many
astrophysical plasmas fed the well-known two-stream (electrostatic or Weibel) instabilities [\cite{H75}].

Even for relativistic temperatures the collisionless (Vlasov) plasma condition applies to
many astrophysical scenarios where dissipation is dominated by wave-particle interactions
rather than binary collisions. In these cases a fundamental kinetic description is required,
but the existing standard nonrelativistic results could be improved by the covariant relativistic
approaches which reproduce the general and relativistically correct dispersion relations in
any frame that is not necessarily inertial.

In the first earlier treatment [\cite{LM03}] (named as Paper I in the next) the nonlinear
three-waves interaction have been investigated in the case of relativistic Vlasov
plasma with isotropic distribution function of particles. Using the relativistic Vlasov-Maxwell
equations the new relativistic forms of nonlinear current and nonlinear plasma conductivity
have been derived. The nonlinear current is the typical nonlinear response of a plasma
medium where a pump and an idler wave interact exciting a third signal wave.

The nonlinear current seems to play an important role also in the
saturation stage of plasma filamentation arising in laser plasma
interaction [\cite{UT02}] or in the jet-plasma astrophysical 
structures [\cite{N03}]: shocks
in GRBs sources, pulsar outflows, or solar winds. The
filamentation instability is the final stage of nonlinear waves
interaction when the plasma distribution function is anisotropic
and which persists as long as the external perturbation is
sufficiently strong. This case should be analyzed using an
anisotropic bi-Maxwellian distribution function to calculate the
nonlinear conductivity, and it will be discused in the next papers
of this series.

In the present paper we continue to develop a fully relativistic
and covariant kinetic theory of nonlinear plasma waves interaction
providing an analytical model to estimate the plasma nonlinear
response. Starting from the Maxwell equations we show in Section
II that two waves coupling leads to a nonlinear component of plasma
current which generates the third wave signal. To find the
corresponding nonlinear fluctuation of plasma density we use as in
Paper I the relativistic nonlinear Vlasov equation to
obtain the general form of the nonlinear current arising from
the interaction of two waves with any polarization and any direction
of propagation.

Considering only longitudinal waves the general covariant expression of nonlinear
plasma conductivity is derived in Section III for any subluminal or superluminal
phase velocity and for the whole complex frequency plane. The nonlinear conductivity
is then calculated in Section IV using an appropriate relativistic distribution function which
is vanishing for particle speeds greater than speed of light. Knowing the nonlinear
conductivity we could find the electric fields of interacting waves as solutions of
nonlinear (coupled) equations system (42)-(44) from Paper I.

For nonrelativistic plasma temperatures we derive in Section V
apparently for the first time, the covariant expression of
nonlinear conductivity in terms of well documented plasma
dispersion function [\cite{fc61}]. We show that the covariant form
is markedly different from the standard classic expression of
nonlinear conductivity provided by the noncovariant
nonrelativistic theory and which can be obtained only in the
unphysical limit of infinite speed of light $c \to \infty$
(because the classical theory assumes all waves as being
subluminal).


\section{BASIC FRAMEWORK}

The basic model followed in this paper is an infinitely extended
collisionless plasma of high temperature electrons which requires
that the electrons be treated relativistically. For plasmas with
low collisionality ($\nu_c/\omega_p \simeq g<<1$), the cooperative
motion is due to the electromagnetic coupling of the particles,
not to collisions. Therefore, the relativistic Vlasov-Maxwell
equations are used, and for plasma fluctuations with sufficiently
large amplitudes, the nonlinear forms of them are required (see
notations and equations (13)-(17) from Paper I).

The equation governing wave propagation in plasmas, is:
\begin{eqnarray}
\left[\nabla \times(\nabla \times) + \frac{1}{c^2}\frac{\partial^2}{\partial
t^2}\right] {\bf E}({\bf r}, t)=-\frac{4\pi}{c^2}\frac{\partial {\bf J}({\bf r},
t)}{\partial t}, \label{eq:ecunda}
\end{eqnarray}
which follows directly from the Maxwell equations (see equations (14)-(17) from
\cite{LM03}). Waves interaction gives rise to the nonlinear terms in ${\bf J}$:
\begin{eqnarray}
{\bf J}({\bf r}, t)={\bf J}^{\rm L}({\bf r}, t)+{\bf J}^{\rm NL}({\bf r}, t),
\label{eq:jtot}
\end{eqnarray}
with
\begin{eqnarray}
{\bf J}^{\rm L}({\bf r}, t)=\sum_i {\bf J}^{\rm L}_i({\bf k}_i, \omega_i) = \sum_i
\sigma(\omega_i){\bf E}_i ({\bf k}_i, \omega_i), \label{eq:jlin}
\end{eqnarray}
and
\begin{eqnarray}
{\bf J}^{\rm NL}({\bf r}, t)=\sum_{p\geq 2} {\bf J}^{\rm NL}_p({\bf r}, t) = \sum_u
{\bf J}^{\rm NL}_u({\bf k}_u, \omega_u)=\sum_u {\bf J}^{0,{\rm NL}}_u e^{-i(\omega_u
t-{\bf k}_u\cdot{\bf r})}. \label{eq:jnelin}
\end{eqnarray}

We assume that
\begin{eqnarray}
{\bf E}({\bf r}, t)&=&\sum_i {\bf E}_i({\bf k}_i, \omega_i) = \sum_i {\bf E}^0_i
e^{-i(\omega_i t-{\bf k}_i\cdot{\bf r})}, \label{eq:electric}
\\
{\bf D}({\bf r}, t)&=&\epsilon {\bf E}({\bf r}, t)=\sum_i \epsilon(\omega_i)
{\bf E}_i({\bf k}_i, \omega_i), \label{eq:delectric}
\end{eqnarray}
and ${\bf E}^0_i$ is taken as essentially independent of position in space.

For the lowest-order nonlinear phenomena ($p=2$), as typical three-waves
configuration, two waves (indices 1 and 2) give rise to the third one (index 3)
through a first generated nonlinear current:
\begin{eqnarray}
{\bf J}_3^{\rm NL}({\bf r}, t)={\bf J}_{12}^{\rm NL}({\bf r}, t) +{\bf J}_{21}^{\rm
NL}({\bf r}, t). \label{eq:jnelin12}
\end{eqnarray}
With (\ref{eq:jlin})-(\ref{eq:electric}) and
\begin{eqnarray}
\epsilon (\omega_i) \equiv 1+\frac{4\pi i}{\omega_i}\sigma(\omega_i),
\label{eq:epsilon}
\end{eqnarray}
(\ref{eq:ecunda}) becomes:
\begin{eqnarray}
\left[\nabla \times(\nabla \times) +\epsilon\,\, \frac{1}{c^2}\frac{\partial^2}{\partial t^2}
\right]{\bf E}({\bf r}, t)=-\frac{4\pi}{c^2}\frac{\partial {\bf J}^{\rm NL}({\bf r},
t)}{\partial t}, \label{eq:ecunda1}
\end{eqnarray}
clearly indicating (with (\ref{eq:electric}) and (\ref{eq:jnelin12})) that wave
electric fields ${\bf E}_i({\bf k}_i, \omega_i)$ are coupled through nonlinear
current ${\bf J}^{\rm NL}$.


\subsection{Nonlinear current}

The nonlinear density fluctuation $f^{\rm NL}$ is solution of the
nonlinear Vlasov equation (see in Ref. [\cite{LM03}]) and it will
determine the nonlinear current:

\begin{eqnarray}
{\bf J}^{\rm NL}_3 &=& {\bf \sigma}^{\rm NL}{\bf E}{\bf E}=q \int_{-\infty}^{\infty}
\, d^3p {\bf v} f^{\rm NL} {}\nonumber \\{}&=& -iq^2 \int_{-\infty}^{\infty} d^3p
\frac{{\bf p}}{\gamma m \omega_3 - {\bf k}_3 \cdot {\bf p}}\left[ \left( {\bf E}_1+
\frac{{\bf p}\times{\bf B}_1}{\gamma m c}\right)\cdot \frac{\partial f_2}{\partial
{\bf p}} + (1 \leftrightarrow 2)\right], \label{eq:jnl}
\end{eqnarray}

where $(1 \leftrightarrow 2)$ denotes the interchanging of indices 1 and
2 required by (\ref{eq:jnelin12}), and
$\gamma = \sqrt{1+p^2 / m^2c^2}$ is the Lorentz factor.
Otherwise, the linear fluctuations
of the distribution function are solutions of the linear Vlasov equation
as
\begin{eqnarray}
f_i = -i\frac{q}{\omega_i R_i}\left( R_i {\bf E}_i + \frac{{\bf p}\cdot {\bf E}_i}
{\gamma m \omega_i}{\bf k}_i \right)\cdot \frac{\partial f_0}{\partial {\bf p}}=
-i\frac{q}{\omega_i R_i} \frac{\partial f_0}{\partial p}\,
\frac{{\bf p}\cdot{\bf E}_i}{p},
\label{eq:f2}
\end{eqnarray}
where an isotropic distribution function $f_0({\bf p})=f_0(p)$, has been assumed at
equilibrium, and the relativistic resonance factors are given by

\be
R_i=1 - ({\bf
p}\cdot{\bf k}_i /\Gamma m \omega_i)\;\;\; (i=1, 2, 3). \label{Ri}
\ee

From (\ref{eq:f2})

\bdm
\frac{\partial f_i}{\partial {\bf p}} = -i\frac{q}{\omega_i R_i p}
\frac{\partial f_0}{\partial p}\,
\left\{\frac{{\bf p} \cdot {\bf E}_i}{p} \left[ \left(p\,\frac{\partial^2 f_0/\partial p^2}
{\partial f_0/\partial p} + \frac{1-R_i -\gamma^2}{\gamma^2 R_i}\right)
\frac{{\bf p}}{p} + \right. \right. \edm 
\be \left. \left.\frac{1- R_i}{R_i} \,\frac{p {\bf k}_i}
{{\bf p}\cdot{\bf k}_i } \right] +{\bf E}_i \right\},\nonumber \\ \label{eq:dfj}
\ee
and a new expression for density current (\ref{eq:jnl}) is derived
\\

${\bf J}^{\rm NL}_3 ={\bf \sigma}^{\rm NL}{\bf E}_1{\bf E}_2=$
\be
-\frac{2 \pi q^3}{m} \int_{-\infty}^{\infty} \, dp_{\parallel} \,
\int_{0}^{\infty} \, dp_{\perp} \,
\frac{{\bf p}\, p_{\perp}}{\gamma \,p}\, \frac{\partial f_0}{\partial
p}\,\frac{{\bf \Omega}}{\omega_1 \omega_2 \omega_3 R_1 R_2 R_3}{\bf E}_1{\bf E}_2,
\label{eq:jnlEab}
\ee
with the second order tensor

\begin{eqnarray}
{\bf \Omega} &=&\left[ \left(p\,\frac{\partial^2 f_0/\partial p^2}
{\partial f_0/\partial p} - \frac{1}{\gamma^2} \right)
(\omega_1R_1+\omega_2R_2) \right. {}\nonumber
\\
{}&&\left.+ \left(1-\frac{1}{\gamma^2}\right)
\left( \frac{({\bf k}_1 \cdot{\bf k}_2)\, c^2}{\omega_1 \omega_2}-1\right)
(\frac{\omega_1R_1}{R_2}+\frac{\omega_2R_2}{R_1})\right]
\frac{{\bf p}{\bf p}}{p^2} {}\nonumber
\\
{}&&+ (\omega_1R_1^2+\omega_2R_2^2)
\left[{\bf I} -\left(1-\frac{1}{R_1}\right) \,\frac{{\bf p}{\bf k}_1}
{{\bf p}\cdot{\bf k}_1 } -\left(1-\frac{1}{R_2}\right) \,\frac{{\bf k}_2 {\bf p}}
{{\bf p}\cdot{\bf k}_2 } \right]. \label{eq:T}
\end{eqnarray}
Comparing with Eq. (25) from Paper I describing only the interaction of parallel propagating
waves, ${\bf k}_i \parallel {\bf k}_j$ ($i,j=1,2,3$), the new general expression
(\ref{eq:jnlEab}) (combined with (\ref{eq:T})), provides the nonlinear current
arising from the interaction of two waves with any polarization and any directions of
propagation.

But for the sake of simplicity we consider in the next only the case
of electrostatic (longitudinal) turbulence (${\bf E}_i\parallel {\bf k}_i$, $i =1,2$)
with interaction of parallel propagating waves (along the same direction)
${\bf k}_1 \parallel {\bf k}_2$.


\section{Covariant case}

In the general covariant case it is convenient to transform to the new variables
of integration [\cite{L67}], $y \equiv \ppa /(mc)$ and
$E\equiv \sqrt{1+{{\ppa ^2+\pper ^2}\over {m^2c^2}}}$, which
yields for the corresponding components of nonlinear conductivity tensor in (\ref{eq:jnlEab})

\be
\sigma_{\parallel}^{NL} = -2\pi q^3 m c^2 \int_1^{\infty} {dE \over E}
{\partial f_0 \over \partial E} \int_{-\sqrt{E^2-1}}^{\sqrt{E^2-1}}\, dy\, y
{\Omega_{LL}\over \omega_1 \omega_2 \omega_3 R_1 R_2 R_3}, \label{spar}
\ee

\be
\sigma_{\perp}^{NL} = -2\pi q^3 m c^2 \int_1^{\infty} {dE \over E}
{\partial f_0 \over \partial E} \int_{-\sqrt{E^2-1}}^{\sqrt{E^2-1}}\, dy\,
\sqrt{E^2-1-y^2} {\Omega_{LL} \over \omega_1 \omega_2 \omega_3 R_1 R_2 R_3}. \label{sperp}
\ee
where $\Omega_{LL}$ is the component of second order tensor ${\bf \Omega}$
corresponding to the interaction of two longitudinal waves
(${\bf E}_i\parallel {\bf k}_i$, $i =1,2$):

\be
\Omm_{LL}= {y^2 \over E}\, \left({\partial^2 f_0\over \partial E^2}/
{\partial f_0\over \partial E}\right) \, (\omm_1 R_1+\omm_2 R_2)+
\left(1-{y^2\over E^2}\right) \, \left({\omm_1 R_1 \over R_2}+
{\omm_2 R_2 \over R_1}\right)
\label{oll}
\ee
and

\be
R_i = 1- {y k_i c \over \omm_i E} = 1- {y \over z_i E}
\;\;\; (i=1, 2, 3).\label{Ril}
\ee
In (\ref{Ril}) the inverse index of refraction $z=\omm /kc$ is introduced.


\subsection{Reduction of $\sigma_{\parallel}^{NL}$}

Introducing (\ref{oll}) in (\ref{spar}) we obtain

\be
\sigma_{\parallel}^{NL} = -2\pi q^3 m c^2 (I_1 +I_2) \label{s12}
\ee
with

\be
I_1 = \int_1^{\infty} \, {dE \over E^2} \, {\partial^2 f_0 \over \partial E^2}
\int_{-\sqrt{E^2-1}}^{\sqrt{E^2-1}}\, dy \, y^3 {1\over \omm_1 \omm_3 R_1 R_3}
+(1 \leftrightarrow 2) \label{i1}
\ee

\be
I_2 = \int_1^{\infty} \, {dE \over E} \,  {\partial f_0 \over \partial E}
\int_{-\sqrt{E^2-1}}^{\sqrt{E^2-1}}\, dy \, y \left(1-{y^2 \over E^2} \right)
{1\over \omm_1 \omm_3 R_1^2 R_3} +(1 \leftrightarrow 2) \label{i2}
\ee
where $(1 \leftrightarrow 2)$ denotes the interchanging of indices 1 and 2.
In terms of the functions $u = u(E)$ and $v = v(E)$ defined by

\be
{\partial u \over \partial E}= -E^2 {\partial^2 f_0 \over \partial E^2},
\;\;\; {\partial v \over \partial E}= -E {\partial f_0 \over \partial E}
\label{uv}
\ee
and using the substitution $t = y /E$ the integrals (\ref{i1}) and (\ref{i2})
from above read

\bdm
I_1 = - {z_1\over \omm_1 \omm_3} \int_1^{\infty} \, dE \, {\partial u \over \partial E}
\int_{-\sqrt{1-E^{-2}}}^{\sqrt{1-E^{-2}}}\, dt \, {t^3\over (z_1 -t)(z_3-t)}
+(1 \leftrightarrow 2) = \edm
\be {2 z_1 \over \omm_1 \omm_3 (z_3- z_1)} \,
\int_1^{\infty} \, dE \, {u \sqrt{E^2 -1}\over E^2}
\left[{z_1^2\over (z_1^2 -1)E^2+1} -{z_3^2\over (z_3^2 -1)E^2+1} \right]
+(1 \leftrightarrow 2)\label{i1p}
\ee

\bdm
I_2 = - {z_1^2 z_3 \over \omm_1 \omm_3}\int_1^{\infty} \, dE \, {\partial v \over \partial E}
\int_{-\sqrt{1-E^{-2}}}^{\sqrt{1-E^{-2}}}\, dt \, {t(1-t^2)\over (z_1 -t)^2(z_3-t)}
+(1 \leftrightarrow 2)= \edm
\bdm {2 z_1^2 z_3 \over \omm_1 \omm_3 (z_3 - z_1)^2} \,
\int_1^{\infty} \, dE \, {v \sqrt{E^2 -1}\over E^4}
\left\{{2 z_1 (z_3 -z_1) E^2\over [(z_1^2 -1)E^2+1]^2} -{1\over (z_1^2 -1)E^2+1} + \right. \edm
\be \left. {1\over (z_3^2 -1)E^2+1} \right\}
+(1 \leftrightarrow 2) . \label{i2p}
\ee
From (\ref{uv})

\be
u(E) = E {\partial v(E) \over \partial E} - 2v(E), \label{uvp}
\ee
which is used in (\ref{i1p}) to find

\bdm
I_1 = {2 z_1 \over \omm_1 \omm_3 (z_3- z_1)} \, \left\{
\int_1^{\infty} \, dE \, {v \over E^2 \sqrt{E^2 -1}}
\left[{z_3^2\over (z_3^2 -1)E^2+1} - {z_1^2\over (z_1^2 -1)E^2+1}\right] +\right.\edm
\be \left. 2 \int_1^{\infty} \, dE \, {v \sqrt{E^2 -1} \over E^2 }
\left[{z_3^2\over [(z_3^2 -1)E^2+1]^2} - {z_1^2\over [(z_1^2 -1)E^2+1]^2}\right] \right\}
+(1 \leftrightarrow 2)\label{i1f}
\ee
and collecting (\ref{i1f}) and (\ref{i2p}) the component (\ref{s12}) of nonlinear
conductivity finally reads

\bdm
\sigma_{\parallel}^{+,NL} = {4\pi q^3 m c^2 \over \omm_1 \omm_3} \;
{z_1 \over z_3- z_1} \, \times \edm
\bdm
\left\{ \left(1+{z_1 z_3 \over z_3 -z_1} \right)
\int_1^{\infty} \, dE \, {v(E) \sqrt{E^2 -1}\over E^4 }
\left[{1\over (1-z_3^2)E^2 -1} - {1\over (1- z_1^2)E^2-1}\right] -\right.\edm
\be
\left. 2 \int_1^{\infty} \, dE \, {v(E) \sqrt{E^2 -1} \over E^2 }
\left[{z_3^2\over [(1- z_3^2)E^2 -1]^2} -
{z_1^2(1-z_3)\over [(1- z_1^2)E^2 -1]^2}\right] \right\}
+(1 \leftrightarrow 2) \label{s12p}
\ee
with $v(E)$ defined by the isotropic distribution function $f_0 (E)$ in (\ref{uv}).

Relation (\ref{s12p}) is the general covariant form of
longitudinal conductivity plasma response to the interaction
of two (longitudinal) plasma waves.
We can use it to find the expressions of plasma waves solving
the nonlinear equation system (42)-(44) from Paper I.
But relation (\ref{s12p}) is not valid for the whole complex
frequency plane because it has been obtained with assumption of
a positive imaginary part, $\Gamma >0$, of frequency $\omm = \omm_r +
\imath \Gamma$ (see also in Paper I). The index "$+$" in (\ref{s12p}) indicates
this condition.

\subsection{Analytic continuation for subluminal waves}

As we noted, the relation (\ref{s12p}) holds for positive values of the imaginary
part of the frequency $\Gamma =\Im \omega >0$, corresponding to $\Im z>0$. In order to derive the
corresponding dispersion relations for negative values of the imaginary
part of frequency, $\Gamma <0$ (i.e. $\Im z<0$), we have to analytically continue
the integrals in (\ref{s12p}) into the negative imaginary plane of the complex variable $z$.

For superluminal waves generally defined in unmagnetized plasma
by $|z_r |= |v_{phase} /c | > 1$, all the integrals
in (\ref{s12p}) admit no pole inside the
integration interval, and we simply have

\be
\sigma^{-,NL}_{\parallel} = \sigma^{+,NL}_{\parallel}, \label{ssup}
\ee
which also holds for plasma waves with subluminal phase velocities
$|z_r|<1$ but with susceptible growing amplitudes $\Gamma >0$.

For subluminal damped waves with $|z_r|<1$ and $\Gamma < 0$
each of the integrals in (\ref{s12p}) admits the pole
$E_{ci}=1 /\sqrt{1-z_i^2} >1$ inside the integration
interval $[1,\infty]$, and the analytical continuation of
(\ref{s12p}) into negative imaginary plane (negative index) reads
as follows
\\

$ \sigma^{-,NL}_{\parallel} = \sigma^{+,NL}_{\parallel} +$
\bdm
\imath {4\pi^2 q^3 mc^2 \over \omm_1 \omm_3}\, {z_1\over z_3 -z_1}
\left\{v (E_{c3}) z^3_3 \sqrt{1-z_3^2} \left[(1-z_3^2)
\left(1+{z_1 z_3 \over z_3 - z_1}\right) -1\right] \right.
\edm
\be
\left. v (E_{c1}) z^3_1 \sqrt{1-z_1^2} \left[(1-z_1^2)
\left(1+{z_1 z_3 \over z_3 - z_1}\right) -1+z_3 \right]\right\}
+(1 \leftrightarrow 2) . \label{ssub}
\ee

\subsection{Interlude}

Combining now (\ref{s12p}) and (\ref{ssub}) we derive
the general form of the longitudinal conductivity

\bdm \sigma^{NL}_{\parallel} = \sigma^{+,NL}_{\parallel} +
\imath {4\pi^2 q^3 mc^2 \over \omm_1 \omm_3}\, {z_1\over z_3 -z_1} \times \edm
\bdm
\left\{H(1-|\Re z_3|)H(-\Im z_3)\,v (E_{c3}) z^3_3 \sqrt{1-z_3^2} \left[(1-z_3^2)
\left(1+{z_1 z_3 \over z_3 - z_1}\right) -1\right] \right.
\edm
\bdm
\left.- H(1-|\Re z_1|)H(-\Im z_1)\, v (E_{c1}) z^3_1 \sqrt{1-z_1^2} \left[(1-z_1^2)
\left(1+{z_1 z_3 \over z_3 - z_1}\right) -1+z_3 \right]\right\}\edm
\be
+(1 \leftrightarrow 2) . \label{sg}
\ee
holding for the whole complex frequency plane, i.e. for interaction of any growing
or damped waves.
The frequency and the wave-number of signal wave (noted here with index 3) are
unknown in the conductivity expression, but they are usually provided by the
conservation laws

\be
\omega_3=\omega_1\pm \omega_2 , \;\;\;
{\bf k}_3 ={\bf k}_1 \pm {\bf k}_2 . \label{momc}
\ee

\section{Relativistic Maxwellian plasma}

The isotropic equilibrium distribution function $f_0$ is supposed to be
Maxwellian. But a rigorous relativistic analysis requires for an
appropriate relativistic distribution function which is vanishing
for particle speeds greater than speed of light. We therefore consider
the Maxwell-Boltzmann-J\"uttner ditribution function

\be
f_0(p)=C e^{-\mu E}, \label{f0}
\ee
with a normalization constant
\be
C={n_0 \mu \over 4\pi (mc)^3\, K_2(\mu)},\,\, \,\,\,\,\,\, \mu = \frac{m
c^2}{k_B T},\label{cmu}
\ee
so that $\int d^3p \, f_0 =n_0$. $K_\nu (\mu)$ denotes the modified
Bessel function.
We use (\ref{f0}) in the second equation of (\ref{uv}) to find

\be
v = -(E+ {1 \over \mu}) \, C \, e^{-\mu E}, \label{v}
\ee
which can be introduced in (\ref{s12p})

\be
\sigma_{\parallel}^{+,NL} = {4\pi q^3 m c^2 \over \omm_1 \omm_3} \;
{z_1 \over z_3- z_1} \left[\left(1+{z_1 z_3 \over z_3 -z_1} \right)J_1 -
2 J_2 \right] +(1 \leftrightarrow 2) \label{sf}
\ee
to calculate the integrals:
\\

$J_1^+ = $ 
\be - \, C \int_1^{\infty} \, dE \, { e^{-\mu E} \sqrt{E^2 -1}\over E^4 }
\left(E+{1 \over \mu} \right)\,
\left[{1\over (1-z_3^2)E^2 -1} - {1\over (1- z_1^2)E^2-1}\right] \label{j1}
\ee
\\

$J_2^+ =$ 
\be-\, C \int_1^{\infty} \, dE \, {e^{-\mu E} \sqrt{E^2 -1} \over E^2 }
\left(E+{1 \over \mu} \right)\, \left[{z_3^2\over [(1- z_3^2)E^2 -1]^2} -
{z_1^2(1-z_3)\over [(1- z_1^2)E^2 -1]^2}\right]
\label{j2}
\ee

\section{Nonrelativistic thermal plasmas ($\mu >> 1$)}

We now consider the limit of nonrelativistic plasma temperatures $\mu >>1$,
and to lowest order in $\mu^{-1}<<1$

\be K_2(\mu )\simeq \sqrt{\pi \over 2(\mu)}e^{-(\mu)}, \;\;\;
C= ({\mu \over 2 \pi})^{3/2} \, {n_0 \over (mc)^3}\,
e^{\mu}.  \label{k2c}
\ee
and with the substitution $E = \sqrt{1+s^2}$ the first integral (\ref{j1})
becomes:

\be
J_1^+ = -\, ({\mu \over 2 \pi})^{3/2} \, {n_0 \over (mc)^3}\,
e^{\mu} \int_0^{\infty} \, ds \, {s^2 \, e^{-\mu \sqrt{1+s^2}} \over (1+s^2)^2}
\left[{1\over (1-z_3^2)s^2 -z_3^2} - {1\over (1- z_1^2)s^2-z_1^2}\right] \label{j1i}
\ee
\\
Because of the exponential function the main contribution to the integral
for large values of $\mu _a>>1$ comes from small values of
$s<<1$, so that we may approximate $\sqrt{1+s^2}-1\simeq s^2/2$ yielding
\\

$J_1^+ =$
\bdm
 -\, \left({\mu \over 2 \pi}\right)^{3/2} \, {n_0 \over (mc)^3}\, \left\{
{1 \over 1-z_3^2}\,\left[ \int_0^{\infty} \, ds \, e^{-\mu s^2 /2}+{z^2_3 \over 1-z_3^2}
\,\int_0^{\infty} \, ds \, {e^{-\mu s^2 /2} \over s^2 -{z^2_3 \over 1-z_3^2}}
 \right]- \right. \edm 
 \be \left. (3 \leftrightarrow 1)\right\}=
-\, {n_0 \mu \over 2^{3/2} \pi (mc)^3} \,
\left[ {f_3 Z^+(f_3) \over 1-z_3^2}-{f_1 Z^+(f_1) \over 1-z_1^2}+{z_3^2 -z_1^2
\over (1-z_1^2)(1-z_3^2)} \right], \label{j1p}
\ee
where we have used (in the positive imaginary frequency plane)
the plasma dispersion function of Fried and Conte [\cite{fc61}]

\be
Z^+ (f) = \pi^{-1/2} \int_{-\infty}^\infty dx \, {e^{-x^2} \over x-f},
\label{z}
\ee
of the argument

\be
f_i = \sqrt{\mu \over 2} {z_i \over \sqrt{1-z_i^2}}. \label{fi}
\ee
And the second integral (\ref{j2}) is calculated in the same manner
as
\be
J_2^+ = {n_0 \mu \over 2^{3/2} \pi (mc)^3} \,
\left[ {f_3 Z^+(f_3/2) \over 4(1-z_3^2)}-{(1-z_3)f_1 Z^+(f_1/2) \over 4(1-z_1^2)}+{z_3^2 -z_1^2
+z_3(1-z_3^2) \over 2 (1-z_1^2)(1-z_3^2)} \right]. \label{j2p}
\ee
\\

For nonrelativistic plasma temperatures ($\mu \gg 1$) we are entitled to
assume $|f| \gg 1$ for which (see the asymptotic approximation
of plasma dispersion function [\cite{fc61})] $Z^+(f/2) \simeq 2 Z^+(f)$
and after summation of the integrals (\ref{j1p}) and (\ref{j2p})
in (\ref{sf}) we obtain

\be
\sigma_{\parallel}^{+,NL} \simeq -{n_0 q^3 \mu \over \sqrt{2} m^2 c \, \omm_1 \omm_3} \;
{z_1 \over z_3- z_1} \left(1+{z_1 z_3 \over z_3 -z_1} \right)J^+
+(1 \leftrightarrow 2) \label{sfnr}
\ee
with $J^+$ given by

\be
J^+ = {f_3 Z^+(f_3) \over 1-z_3^2}-{(1+z_3)f_1 Z^+(f_1) \over 1-z_1^2}+
{z_3^2 -z_1^2 -z_3(1-z_3^2) \over (1-z_1^2)(1-z_3^2)}. \label{j}
\ee

For negative imaginary frequencies we can find
$\sigma_{\parallel}^{-,NL} $ substituting (\ref{v}) in
(\ref{ssub}), or it is easier to use in (\ref{j}) the well-known
analytic continuation of the plasma dispersion function into the
negative imaginary frequency plane ($\Gamma < 0$) [\cite{r69}]

\be
Z^-(f)={1\over \pi ^{1/2}}\int_{-\infty }^\infty\; dx{e^{-x^2}\over x-f}
+2\imath \sqrt{\pi }e^{-f^2},\,\,\; \Im (f)<0,
\label{z-}
\ee
and the plasma waves are covariantly described now for all complex frequencies by

\be
\sigma_{\parallel}^{NL} = -{n_0 q^3 \mu \over \sqrt{2} m^2 c \,\omm_1 \omm_3} \;
{z_1 \over z_3- z_1} \left(1+{z_1 z_3 \over z_3 -z_1} \right)J
+(1 \leftrightarrow 2) \label{sf-nr}
\ee
where $J$ is given by

\be
J= J^+ +2 \imath \sqrt{\pi } \left[ H(-\Im z_3){f_3 \over 1-z_3^2}e^{-f_3^2}
-H(-\Im z_1){(1+z_3)\, f_1 \over 1-z_1^2}e^{-f_1^2}\right]. \label{j-}
\ee


\subsection{Limit infinite speed of light $c \to \infty$}

We finaly consider the plasma waves interaction
in the formal limit of an infinitely large speed of light, $c\to \infty$,
which corresponds to the classical noncovariant nonrelativistic theory and
where all waves have subluminal phase velocities.
From Eq. (\ref{sfnr})--(\ref{j-}) we obtain in this limit

\bdm
\sigma_{\parallel}^{NL, \infty} = - \sqrt{2}\, {n_0 q^3 c \over  m^2 v^2_{th}}
\,{1\over \omm_1 \omm_3 } \left({k_1\over \omm_1}\,{\omm_3 \over k_3} -1 \right)^{-1}\;
\left[ f^{\infty}_3 \left(Z^+(f^{\infty}_3) + \right. \right. \edm
\be \left. \left. 2 \imath \sqrt{\pi} H(-\Im \omm_3)
e^{-(f^{\infty}_3)^2}\right)
-f^{\infty}_1 \left(Z^+(f^{\infty}_1) + 2 \imath \sqrt{\pi} H(-\Im \omm_1)
e^{-(f^{\infty}_1)^2}\right)\right]
+(1 \leftrightarrow 2) \label{sfnri}
\ee
with

\be
v^2_{th} = {2k_B T \over m}, \;\;\; f^{\infty}_i = {\omm_i \over k_i v_{th}}. \label{fii}
\ee
and which holds for all complex frequencies and for all wave-numbers.

Assuming again $|f^{\infty}| \gg 1$, the exponential functions from
the residues in (\ref{sfnri}) are very small ($e^{-(f^{\infty})^2} \ll 1$)
and the asymptotic approximation of plasma dispersion function in (\ref{sfnri}) leads to

\be
\sigma_{\parallel}^{NL, \infty} \simeq
- \, {n_0 q^3 \, N_3 (N_1 +N_3) \over \sqrt{2} m^2 c\, \omm_1 \omm_3 }
+(1 \leftrightarrow 2),\label{sfnria}
\ee
which agrees with the standard results [\cite{WW77}]
and where $N = kc / \omm$ is the index of refraction.

\section{CONCLUSION}

On the basis of relativistic Vlasov-Maxwell equations we have
developed a covariant kinetic formalism to determine the nonlinear
plasma conductivity which allows us to find all the plasma waves
nonlinearly excited in plasma. The general covariant form of
nonlinear conductivity is provided first for any value of plasma
temperature and for the whole complex frequency plane by a correct
analytical continuation. Then, we have restricted the analysis to
an appropriate relativistic particle distribution which is
vanishing for particle speeds greater than speed of light. And in
the limit of nonrelativistic plasma temperatures we have derived
apparently for the first time the covariant nonlinear conductivity
which is significantly different from the standard noncovariant
nonrelativistic results. Only in the strictly unphysical formal
limit of an infinitely large speed of light $c \to \infty$ the
covariant forms reduce to the standard expressions.

\begin{acknowledgments}
M.L. is grateful to the Alexander von Humboldt Foundation for
supporting in part this work. This work was partially supported
by the Deutsche Forschungsgemeinschaft
through the Sonderforschungsbereich 591.
\end{acknowledgments}

\end{document}